\documentstyle[prd,aps,preprint,floats,graphicx, axodraw]{revtex}

\tightenlines

\begin{document}

\preprint{\vbox{
\hbox{\bf UMD-PP-04-010}
}}

\title{{\Large\bf  Proton Decay in a Minimal SUSY SO(10) Model for
Neutrino Mixings}}

\author{\bf H. S. Goh, R.N. Mohapatra, S. Nasri and Siew-Phang Ng }

\address{Department of Physics, University of Maryland, College
Park, MD 20742, USA}
\date{November, 2003}

\maketitle

\begin{abstract}
A minimal renormalizable SUSY SO(10) model with B-L symmetry broken by
{\bf 126} Higgs
field has recently been shown to predict all neutrino mixings and the
ratio $\Delta m^2_{\odot}/\Delta m^2_A$ in agreement with
observations. Unlike models where B-L is broken by {\bf 16} Higgs, this
model guarantees automatic R-parity conservation and hence a stable dark
matter as well as the absence of dim=4 baryon violating operator without
any additional symmetry assumptions. In this paper, we
discuss the predictions of the model for proton decay induced at the
GUT scale. We scan over the parameter space of the model allowed by
neutrino data and find upper bounds on the partial lifetime for the modes
$\tau(n\rightarrow \pi^0\bar{\nu})=~2\tau(p\rightarrow \pi^+\bar{\nu})\leq
 (5.7-13)\times 10^{32}$ yrs and $\tau(n\rightarrow K^0\bar{\nu})\leq
2.97\times 10^{33}$ yrs for the
average squark mass of a TeV and wino mass of 200 GeV, when
 the parameters satisfy the present lower limits on
$\tau(p\rightarrow K^+\bar{\nu})$ mode. These results can be used to test
the model.
 \end{abstract}

\section{Introduction}
 Grand unified theories\cite{pati} provide an attractive framework for
understanding the origin of the diverse strengths of the various
forces observed in Nature. The basic idea is to have a
single force associated with a grand unified local symmetry at a
high scale, which below the scale of the symmetry breaking evolves
into three different strengths corresponding to the observed weak,
electromagnetic and strong interactions. The challenge is to have a
theory where the three couplings evolved down to the Z-boson mass
scale match their experimentally observed values. A concrete
realization of this hypothesis is provided within the framework of
supersymmetric models where the unification scale is $M_U\simeq
2\times 10^{16}$ GeV, provided one assumes that there are no
intermediate scales below $M_U$ and the scale of supersymmetry
breaking is at the weak scale\cite{many}. One may then assume that above
the
scale $M_U$, the grand unified symmetry takes over. While this
observation suggests that the idea grand unification may be
correct, one cannot take it as an evidence for it due to the extra
assumption of grand desert which is crucial to obtaining unification
of couplings.

Typical grand unified theories not only unify forces but also
quarks and leptons and thereby provide a hope to resolve some of
the puzzles of the standard model. Especially after the discovery
of neutrino masses and mixings, there are a large number of
observables in the matter sector that are known fairly precisely
and one may expect that a grand unified theory can provide an
explanation of these observables.

The obvious question then is: how can one test the idea of grand
unification and more precisely, the existence of a local grand
unifying symmetry of all matter and forces ? It is well known that
one of the consequences of most grand unified theories is that
proton is unstable and therefore one may argue that the
 true test of the grand unification hypothesis will come from the
observation of proton decay. This has led to many experimental as
well as theoretical investigations of  proton decay in the context
of simple grand unified models such as SU(5) and SO(10). Although
no evidence for proton decay has been found to date, the stringent
experimental upper limits on the partial lifetimes to various
modes have provided new and useful constraints on the nature of
grand unified theories.

The simplest grand unified theory is the minimal SUSY SU(5) model.
In this model, the dominant decay of proton occurs via dimension
five operators involving color triplet Higgsino exchange leading
to the dominant decay mode \cite{raby} $p\rightarrow
K^+\bar{\nu}$. The predictions of the minimal renormalizable SU(5)
model for this mode has been discussed in many
papers\cite{su51,su52}\footnote{By renormalizable, we mean a
theory where only renormalizable terms are included in the
superpotential.}. The present experimental lower limit on this
mode\cite{SK} is $1.9\times 10^{33}$ yrs, which is an order of
magnitude larger than prediction of the minimal renormalizable
SUSY SU(5) model. Therefore this model is ruled out. It has been
shown\cite{su52} that if one includes nonrenormalizable terms in
the superpotential\cite{su53}, one can get somewhat higher
lifetimes for this decay mode and the SU(5) model can still be
consistent with experiments\footnote{For proton decay in string theories
with SU(5) GUT, see \cite{witten}.}.

With the discovery of neutrino masses and mixings\cite{alexei},
SO(10) is a much more interesting candidate for grand unification.
It incorporates the right handed neutrino needed for implementing
the seesaw mechanism\cite{seesaw} for understanding small neutrino
masses. In addition, unlike SU(5), it incorporates all the
fermions of the standard model as well as the right handed
neutrino into one spinor representation. Furthermore, it is
interesting that using the value of the neutrino mass difference
required to understand atmospheric neutrino data and the seesaw
formula, one can conclude that at least one of the right handed
neutrino masses must be close to $10^{15}$ GeV. The proximity of
this scale to $M_U$ suggests that they may indeed be one and the
same. This would imply that seesaw mechanism may be the first
signal of the idea of grand unification. It would then be urgent
to look for other signals for SO(10) grand unification such as
proton decay\cite{raby1}.

It has recently been pointed out that there is a class of minimal
SO(10) models, where all neutrino masses and mixings can be predicted
without assuming any additional symmetries\cite{babu,goran,goh}. The basic
puzzle of neutrino physics i.e. the two large mixings $(\theta_{\odot}<
\theta_A)$ are explained in this model in a very interesting manner. All
the Yukawa couplings that are responsible for proton decay are completely
fixed in this model. Our goal in this paper is to study the proton decay
in this model.

The key ingredient of our model is the way
B-L local symmetry in SO(10) is broken. How B-L is broken decides whether
the effective MSSM theory that emerges below the GUT scale
conserves R-parity symmetry (defined by $(-1)^{3(B-L)+2S}$) or not. As is
well-known, R-parity symmetry is needed to guarantee the existence of a
stable dark matter. The B-L symmetry could either be broken by
the {\bf 16} Higgs or by the {\bf 126} Higgs multiplet.  We are driven to
using the {\bf 126}-Higgs since, unlike breaking by {\bf 16},
{\bf 126} breaks B-L by two units and leaves R-parity unbroken
as can be concluded from the formula for R-parity given above and
hence a stable dark matter\cite{lee}.

The main observation in the papers\cite{babu,goran,goh} is that if B-L
symmetry is broken by a {\bf 126} dimensional representation of
SO(10), the coupling of {\bf 126} to fermions unifies the flavor structure
of the quarks and
charged leptons with those of the neutrinos. This makes the model
remarkably predictive in the neutrino sector and the predictions
obtained in \cite{goh} are now in full accord with all known data for
neutrino mixings. Furthermore, it is
interesting that the experimentally much sought
after parameter $U_{e3}$ is predicted in the model to lie between
0.15-0.18, which can be probed in very
near future in the MINOS experiment at Fermilab as well as several
future experiments being proposed providing a test of the model.

In order to study proton decay in this model, we consider all
dimension five operators\cite{sakai}. There are LLLL as well as
RRRR type operators in this theory. We find that LLLL type
operators dominate proton decay. This part of the discussion is
similar to that of the SU(5) model. However, unlike the minimal
SU(5), in the minimal SO(10) model there are several contributions
which for some domain of parameters can partially cancel each
other. The cancellation is however not complete so that the net
effect is to suppress the decay rate. We are then able to find
upper limits on the proton lifetime. The lifetimes to various
p-decay modes with charged lepton final states can be predicted as
a function of the color triplet Higgsino mass and SUSY breaking
parameters such as the wino mass and squark masses. For a specific
choice of these parameters in the supersymmetry breaking sector (i.e.
average $M_{\tilde{q}}~=~1$ TeV and $M_{gluino}~=~200~ GeV$),
we find upper bounds for the partial lifetimes for the modes
$\tau(n\rightarrow
\pi^0\bar{\nu})=~2\tau(p\rightarrow \pi^+\bar{\nu})\leq
 (5.7-13)\times 10^{32}$ yrs and $\tau(n\rightarrow K^0\bar{\nu})\leq
2.97\times 10^{33}$ yrs
We also give the partial lifetimes for other charged lepton modes for
these cases.

This paper is organized as follows: in sec.2, we review the basic outline
of the model. In sec 3, we discuss the effective operators for the various
proton decay modes, their origin and dependence on the
parameters of the theory.
In the first part of sec.4, we present our predictions for the upper
limits as well as the
allowed ranges for the partial lifetimes to various modes.
 In sec. 5, we present our conclusions.

\section{Brief overview of the minimal SO(10) model with {\bf 126}}
 The SO(10) model that we will work
with in this paper has the following features: It contains three
spinor {\bf 16}-dim. superfields that contain the matter fields
(denoted by $\psi_a$); two Higgs fields, one in the {\bf 126}-dim
representation (denoted by $\Delta$) that breaks the
$SU(2)_R\times U(1)_{B-L}$ symmetry down to $U(1)_Y$ and another
in the {\bf 10}-dim representation ($H$) that breaks the
$SU(2)_L\times U(1)_Y$ down to $U(1)_{em}$. These are the only two
Higgs multiplets that couple to fermions and after symmetry
breaking give rise to all the fermion masses including the
neutrinos. The original SO(10) symmetry can be broken down to the
left-right group $SU(2)_L\times SU(2)_R\times U(1)_{B-L}$ by ${\bf
54}\oplus {\bf 210}$ Higgs fields denoted by $S$ and $\Sigma$
respectively. We wish to point out that the minimal realistic SO(10) model
of the type under discussion can be constructed without including a {\bf
54} Higgs field\cite{minimal}. However we include it here since it
provides the most general description of proton decay.

To see what this model implies for fermion masses, let us explain how the
MSSM doublets emerge and the consequent fermion mass sumrules they
lead to. As noted, the
{\bf 10} and $\overline{\bf 126}$ contain two
(2,2,1) and (2,2,15) submultiplets (under $SU(2)_L\times SU(2)_R\times
SU(4)_c$ subgroup of SO(10)). We denote the two pairs by $\phi_{u,d}$
and $\Delta_{u,d}$\footnote{It must be pointed out that if the
SO(10) symmetry is broken by a {\bf 210} multiplet, then a new Higgs
doublet pair from the (2,2,20)$\oplus$(2,2,$\bar{10}$) multiplets also
mixes with the afore mentioned doublets. But this simply redefines the
mixing angles $\alpha_{u,d}$ and does not affect any of our results}. At
the GUT scale, by some doublet-triplet
splitting mechanism these two pairs reduce to the MSSM Higgs pair
$(H_u,H_d)$, which can be expressed in terms of the $\phi$ and $\Delta$ as
follows:
\begin{eqnarray}
H_u &=& \cos\alpha_u \phi_u + \sin\alpha_u \Delta_u \\
\nonumber
H_d &=& \cos\alpha_d \phi_d + \sin\alpha_d \Delta_d
\end{eqnarray}
The details of the doublet-triplet splitting mechanism that leads to the
above equation are not relevant for what follows and we do not discuss it
here. As in the case of MSSM, we will assume that the Higgs doublets
$H_{u,d}$ have the vevs $<H^0_u>=v \sin\beta$ and $<H^0_d>=v \cos\beta$.

In orders to discuss fermion masses in this model, we start with the
SO(10) invariant superpotential giving the Yukawa couplings of the {\bf
16} dimensional matter spinor $\psi_i$ (where $i,j$ denote generations)
with the Higgs fields $H_{10}\equiv
{\bf 10}$ and $\Delta\equiv {\bf \overline{126}}$.
\begin{eqnarray}
{W}_Y &=&  h_{ij}\psi_i\psi_j H_{10} + f_{ij} \psi_i\psi_j\Delta
\end{eqnarray}
SO(10) invariance implies that $h$ and $f$ are symmetric matrices.
We ignore the effects coming from the higher dimensional operators,
as we mentioned earlier. Also we set all CP phases in the superpotential
as well as vevs to zero, so that the observed kaon and B-CP violation is a
consequence of all CP phases residing in the squark sector.

Below the B-L breaking (seesaw) scale, we can write the superpotential
terms for the charged fermion Yukawa couplings as:
\begin{eqnarray}
W_0 &=& h_u QH_uu^c + h_d QH_d d^c + h_eLH_d e^c + \mu H_uH_d
\end{eqnarray}
where
\begin{eqnarray}
h_u &=& h\cos\alpha_u + f \sin\alpha_u\\ \nonumber
h_d &=& h\cos\alpha_d + f \sin\alpha_d\\ \nonumber
h_e &=& h\cos\alpha_d -3 f \sin\alpha_d
\end{eqnarray}
In general $\alpha_u\neq \alpha_d$ and this difference is responsible for
nonzero CKM mixing
angles. In terms of the GUT scale Yukawa couplings, one can write the
fermion mass matrices (defined as ${\cal L}_m~=~\bar{\psi}_LM\psi_R$) at
the seesaw scale as:
\begin{eqnarray}
M_u &=& \bar{h} + \bar{f} \\ \nonumber
M_d &=& \bar{h}r_1 + \bar{f}r_2 \\ \nonumber
M_e &=& \bar{h}r_1 -3r_2 \bar{f} \\ \nonumber
M_{\nu^D} &=& \bar{h} -3 \bar{f}
\label{sumrule}\end{eqnarray}
where
\begin{eqnarray}
\bar{h} &=& h \cos\alpha_u \sin\beta v_{wk}\\ \nonumber
\bar{f} &=& f \sin\alpha_u \sin\beta v_{wk}\\ \nonumber
r_1 &=& \frac{\cos\alpha_d}{\cos\alpha_u}\cot\beta\\ \nonumber
r_2 &=& \frac{\sin\alpha_d}{\sin\alpha_u}\cot\beta
\end{eqnarray}
The mass sumrules in Eq. (\ref{sumrule}) were crucial to the predictivity
of the model. The neutrino masses are then predicted using the type II
seesaw formula\cite{seesaw2}:
\begin{eqnarray}
{\cal M}_\nu \simeq f\frac{v^2_{wk}}{\lambda
v_{B-L}}-\frac{m^2_D}{fv_{B-L}}.
\label{type2}\end{eqnarray}
 In models which have asymptotic parity
symmetry such as left-right or SO(10) models, it is the type II seesaw
that is more generic than the conventional seesaw formula. For some
parameter range, the first term may dominate. What is important for our
considerations is that the matrix $f$ that determines the flavor structure
of the neutrinos is related to the quark and lepton masses\footnote{In
extensions of the standard model with triplets, one has only the first
term in the neutrino mass formula\cite{valle}. In these models however,
the flavor structure of the neutrinos is completely independent of the
quark sector.}.

Using Eq.\ref{sumrule} and \ref{type2}, in reference \cite{goh} the
neutrino masses and mixings have been calculated. We do not display those
predictions here. But we note that by scanning over the allowed values for
the extrapolated quark and lepton masses as well as quark mixings, we find
a range of predictions for the neutrino sector. We find that a large range
of the predictions are disfavored by the latest solar
data\cite{SNO}. However, there is also a significant allowed region. For
this region, we extract all the Yukawa parameters $\bar{h}_{ij}$ and
$\bar{f}_{ij}$ corresponding to this range and use them in our calculation
of proton decay rate below. A typical set of values for $h$'s and $f$'s
in this range are:
\begin{eqnarray}
h~=~\pmatrix{3.26\times 10^{-6} & 1.50\times 10^{-4} & 5.51\times
10^{-3} \cr 1.50\times 10^{-4} & -2.40\times 10^{-4} & -0.0178 \cr
5.51\times 10^{-3} & -0.0178 & 0.473}
\end{eqnarray}
and
\begin{eqnarray}
f~=~\pmatrix{-7.04\times 10^{-5} & -2.05\times 10^{-5} & -7.53
\times 10^{-4} \cr  -2.05\times 10^{-5} & -1.85\times 10^{-3} &
2.43\times 10^{-3} \cr -7.53 \times 10^{-4} & 2.43\times 10^{-3} &
-1.64\times 10^{-3} }.
\end{eqnarray}

\section{Effective operators for proton decay}
In our model, there are four supersymmetric graphs that contribute to
$\Delta B=1$ operator. They are given in Fig. 1 and involve the exchange
of {\bf 10}, $\overline{\bf 126}$\cite{nath1} Higgs multiplets and two
mixed $10-126$
diagrams. They will lead to both LLLL as well as RRRR type contributions
given by the following effective superpotential:
\begin{eqnarray}
{\cal W}_{\Delta B=1}~=~ M^{-1}_T[C_{ijkl}
\epsilon_{\alpha\beta\gamma}Q^{\alpha}_i
Q^{\beta}_jQ^{\gamma}_k L_l~+~D_{ijkl}
\epsilon_{\alpha\beta\gamma}u^{c,\alpha}_i
d^{c,\beta}_ju^{c,\gamma}_k e^c_l]
\end{eqnarray}
where $M_T$ is the effective  mass of color triplet field.
\begin{figure}[t]
\begin{center}
\begin{picture}(400,220)(0,0)

\SetOffset(100,0)
 \ArrowLine(20,10)(60,50)
  \ArrowLine(20,90)(60,50)
   \ArrowLine(180,10)(140,50)
    \ArrowLine(180,90)(140,50)
\Line(60,50)(140,50) \Line(100,50)(100,80) \GCirc(100,84){5}{0.5}

\Text(100,50)[c]{$\times$} \Text(30,5)[c]{$16_M$}
\Text(30,95)[c]{$16_M$} \Text(170,95)[c]{$16_M$}
\Text(170,5)[c]{$16_M$} \Text(80,58)[c]{$\overline{126}_H$}
\Text(120,58)[c]{$10_H$} \Text(100,95)[c]{$210$} \Text(100,
15)[c]{$(c)$}

\SetOffset(0,120)
 \ArrowLine(20,10)(60,50)
  \ArrowLine(20,90)(60,50)
   \ArrowLine(180,10)(140,50)
    \ArrowLine(180,90)(140,50)
\Line(60,50)(140,50) \Line(100,50)(100,80) \GCirc(100,84){5}{0.5}

\Text(100,50)[c]{$\times$} \Text(30,5)[c]{$16_M$}
\Text(30,95)[c]{$16_M$} \Text(170,95)[c]{$16_M$}
\Text(170,5)[c]{$16_M$} \Text(80,58)[c]{$10_H$}
\Text(120,58)[c]{$10_H$}  \Text(100,95)[c]{$54$}
\Text(100,15)[c]{$(a)$}

\SetOffset(200,120)
 \ArrowLine(20,10)(60,50)
  \ArrowLine(20,90)(60,50)
   \ArrowLine(180,10)(140,50)
    \ArrowLine(180,90)(140,50)
\Line(60,50)(140,50) \Line(100,50)(100,80) \GCirc(100,84){5}{0.5}

\Text(100,50)[c]{$\times$}\Text(30,5)[c]{$16_M$}
\Text(30,95)[c]{$16_M$} \Text(170,95)[c]{$16_M$}
\Text(170,5)[c]{$16_M$} \Text(80,58)[c]{$\overline{126}_H$}
\Text(120,58)[c]{$\overline{126}_H$} \Text(100,95)[c]{$54$}
\Text(100, 15)[c]{$(b)$}

\end{picture}
\caption{\label{Fig1} Superfield Feynman graphs that give rise to
$d=5$ effective proton decay operators.}
\end{center}
\end{figure}
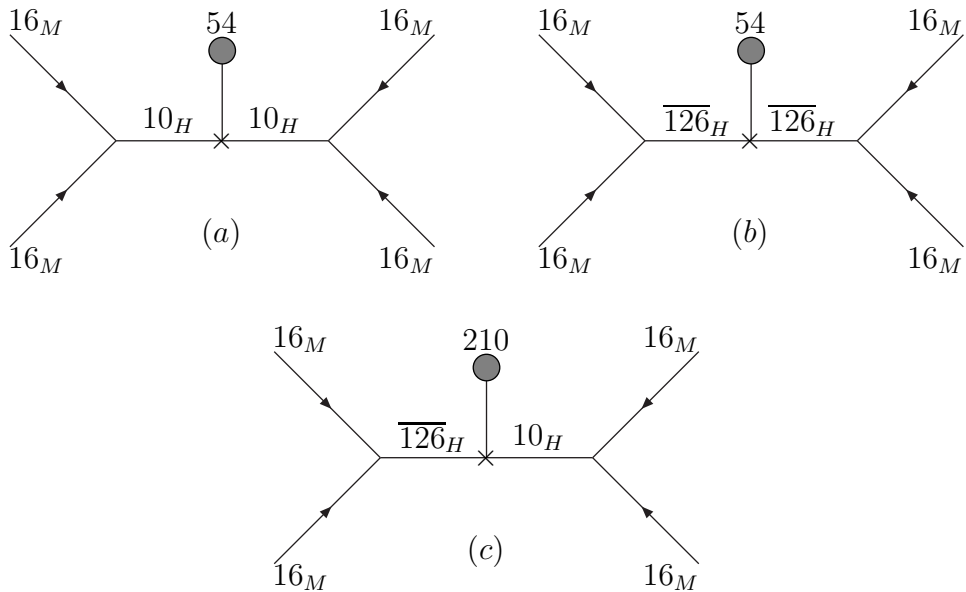

 Note that one
could in principle, diagonalize the mass matrix involving the
color triplet superfields and write the Feynman diagrams in that
basis. It is not hard to convince one self that the final result
in this case will also have four parameters- an effective mass and
three products of mixing angles. So by considering the above
parametrization, we have not lost any information. This
supersymmetric operator leads to effective dimension five
operators that involve two quark (or quark-lepton) fields and two
superpartner fields. In order for these operators to lead to a
Four Fermi operator for proton decay, they must be ``dressed'' via
the exchange of gluinos, winos, binos etc. Before we discuss this,
let us first note that these operators must be antisymmetrized in
flavor indices and then we get for the LLLL term
\begin{eqnarray}
{\cal W}_{\Delta B=1}~=~\epsilon_{\alpha\beta\gamma}
M^{-1}_T[(C_{ijkl}-C_{kjil})u^\alpha_id^\beta_ju^\gamma_ke_l~-
(C_{ijkl}-C_{ikjl})u^\alpha_id^\beta_jd^\gamma_k\nu_l
\end{eqnarray}
There is a similar operator for the RRRR terms. As has been argued
by various authors\cite{su50,nihei}, for small to moderate $\tan
\beta$ region of the supersymmetry parameter space, these
contributions are smaller than the LLLL contributions. We also
find this to be the case in our model. We will show this later;
for the time being therefore, we will focus on the LLLL operator.

The effective four fermion operator responsible for proton decay can arise
 the gluino, bino and wino dressing of the above operators.
The coefficient $C_{ijkl}$ associated with the LLLL terms is expressible
in
terms of the products of the Yukawa couplings $h$ and $f$ which have
already been determined by the neutrino and other fermion masses:
\begin{eqnarray}
C_{ijkl}~=~h_{ij}h_{kl}+xf_{ij}f_{kl}+yh_{ij}f_{kl}+zf_{ij}h_{kl}
\end{eqnarray}
where $x,y,z$ are the ratios of the color triplet masses and mixings. As
already noted, we
do not need to know the detailed form for these parameters ($x,y,z$) in
terms of these masses and mixings. In the end we will vary these
parameters to get the maximum value for the partial lifetimes for the
various decay modes.

We now discuss the dressing of the various terms. The typical diagrams are
shown in Fig.2.

\begin{figure}[t]
\begin{center}
\begin{picture}(190,100)(0,0)

 \ArrowLine(70,50)(20,10)
 \ArrowLine(20,90)(70,50)
 \DashArrowLine(120,90)(70,50){1}
  \DashArrowLine(120,10)(70,50){1}
  \Vertex(70,50){2}
\Photon(120,90)(120,10){3}{7.5} \Line(120,90)(120,10)

 \ArrowLine(170,90)(120,90)
 \ArrowLine(170,10)(120,10)
\Text(15,10)[c]{$f$} \Text(15,90)[c]{$f$} \Text(175,90)[c]{$f$}
\Text(175,10)[c]{$f$} \Text(135,50)[c]{$\widetilde{W}$}
\Text(100,15)[c]{$\widetilde{f}$}
\Text(100,85)[c]{$\widetilde{f}$}

\end{picture}
\caption{\label{Fig2} Generic Feynman graph for dressing of $d=5$
effective proton decay operators via gluino, Wino, Bino and Higgsinos.}
\end{center}
\end{figure}
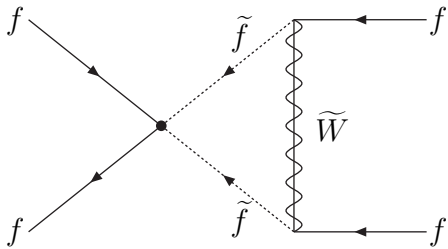

\subsection{Gluino dressing}
It has been pointed out in several papers\cite{vysotsky} in the
limit of all squark masses being same as in mSUGRA type models,
these contributions to the effective four-Fermi operator for
proton decay vanishes. It results from the use of Fierz identity
for two component spinors which states
\begin{equation}
 (\phi_1\phi_2)(\phi_3\phi_4)+(\phi_1\phi_3)(\phi_2\phi_4)+
(\phi_1\phi_4)(\phi_2\phi_3)=0
\end{equation}
$\phi_i$ are the chiral two component spinors representing quarks and
leptons and $(AB)=A^{\alpha}B_{\alpha}\equiv
\epsilon^{\alpha\beta}A_{\alpha}B_{\beta}$
where $\alpha$ and $\beta$ are the spinor indices
($\alpha,\beta=1,2$). Since
satisfying the flavor changing neutral current (FCNC) constraints
allow only very small deviations from universality of squark masses, the
gluino diagrams should be small (proportional to $\delta_{LL,ij}$ in
standard notation\cite{masiero}) in realistic models. We will therefore
ignore these contributions.

The same results hold also for the RRRR operators.

\subsection{Neutral Wino and Bino contribution}

 To analyze the contribution from $\widetilde{W}^o$ and
$\widetilde{B}$, we choose the the operator $\Omega_e=U_i^\alpha
D_j^\beta U_k^\gamma E_l$ as an example. Note that we can use
$\widetilde{W}^o$ and $\widetilde{B}$ in the loop instead of the
the superpartners of Z boson  and photon is because they are both
mass eigenstates due the the assumption of the universal mass.

\subsubsection{$\widetilde{B}$ dressing}
There are 6 different dressings of the operator $\Omega_e$ through
$\widetilde{B}$. We can split them into two groups. One group
involves the lepton and the other one does not. Within each group,
the product the hypercharges from the two vertices are same. Each
of these groups then gives zero due to the Fierz identity as in
the case of gluino dressing. This show that the $\widetilde{B}$
dressing is zero by the same Fierzing argument as the gluino case
in the limit of universal squark masses.

\subsubsection{$\widetilde{W}^o$ dressing}

For the $\widetilde{W}^o$ case, the vertex involving the lepton is
same as that of quarks but different by  a negative sign between
up type and down type particles. The dressing of uu and  dd are
then different from that of ud by a negative sign. Because the
$\widetilde{W}^o$ are lepton/quark blind and the dressing does not
change anything except from boson to fermion, the two groups we
used in the $\widetilde{B}$ analysis are the same. So after
dressing, we have
\begin{equation}
\Omega_e \rightarrow 2(-(u_i^\alpha d_j^\beta)( u_k^\gamma
e_l)-(u_i^\alpha e_l)( u_k^\gamma d_j^\beta)+(u_i^\alpha
u_k^\gamma )(d_j^\beta e_l))
\end{equation}
By the Fierz identity, the sum of the first two terms is equal to
the third and so we have
\begin{equation}
\Omega_e \rightarrow 4(u_i^\alpha u_k^\gamma )(d_j^\beta e_l)
\end{equation}

Due to the antisymmetry of this expression in the color indices,
it is antisymmetric in the interchange of $i$ and $j$. This
implies that $i$ must be different from $k$ and so the two up
quarks belong to different family. This antisymmetry remains true
even after we pass to the mass eigenstate basis, as is easily
checked. The result is simply due to $(u_i^\alpha u_k^\gamma)
=-(u_k^\alpha u_i^\gamma) $. The conclusion is that there is no
$K^0+e^+_l$ or $\pi^o+e^+_l$ decay mode from the $\widetilde{W}^o$
dressing. For the same analysis, the operator $U_i^\alpha
D_j^\beta D_k^\gamma \nu_l$ gives $4(d_j^\beta
d_k^\gamma)(u_i^\alpha\nu_l)$ and so it only contributes to $K^+ +
\bar{\nu}_l$ decay mode.

\subsection{Wino contribution}
In view of the discussion just given the dominant contribution to proton
decay arises from charged wino exchange converting the two sfermions to
fermions.
These diagrams have been evaluated in earlier
works\cite{su51,su52}; we will assume that all scalar superpartners
have the same mass. This leads to the following effective Hamiltonian:
\begin{eqnarray}
{\cal L}_{\Delta
B=1}~=~2I\epsilon_{\alpha\beta\gamma}(C_{kjil}-C_{ijkl})[u^{\alpha,T}_k
Cd^{\beta}_jd^{\gamma,T}_iC\nu_l+
u^{\beta,T}_jCd^{\gamma}_ku^{\alpha,T}_iCe_l]\\ \nonumber
\end{eqnarray}
where $I$ is given by
$I~=~\frac{\alpha_2}{4\pi}\frac{m_{\widetilde{W}}}{M^2_{\widetilde{f}}}$
Using this expression and adding a similar contribution from
$\tilde{W}^0$ exchange, we can now write down the $C$ coefficients
for the different proton decay operators. Table I lists the
total contributions to the different operators in the leading order:
\begin{center}

{\bf Table I}

\bigskip

\begin{tabular}{|c||c|}\hline
Operator & $C$-coefficient \\ \hline ${\bf u d d \nu_\ell}$ & $2I
\sin\theta_c (C_{211l}-C_{112l})$ \\ \hline ${\bf u s d \nu_\ell}$
& $2I(C_{112l}-C_{121l})$ \\ \hline ${\bf u d s \nu_\ell}$ & $2I
\sin \theta_c (C_{221l}-C_{212l})$ \\ \hline ${\bf u d u e_\ell}$
& $2I \sin \theta_c (C_{211l}-C_{112l})$ \\ \hline ${\bf u s d
e_\ell}$ & $2I(C_{112l}-C_{121l})$ \\ \hline
\end{tabular}
\end{center}

{\bf Table Caption:} The coefficients for various $\Delta B=1$
operators from the GUT theory. The $C$'s are products of the
Yukawa couplings in the superpotential as in Eq. (12).

\subsection{Estimates of the RRRR operators}
In this subsection, we give an estimate of the RRRR operators and
confirm that they are indeed negligible compared to to the LLLL
operator contributions for moderate $\tan \beta$ region that we
are interested in. First we note that the gluino dressing graphs
are zero in the limit of all squark and slepton masses being
equal, by the same argument as for the LLLL operators. Secondly,
since all superfields in this operator are $SU(2)_L$ singlets,
there are no wino contribution to leading order. The only
contributions are therefore from the bino exchange and the
Higgsino exchange.

Bino exchange generates a four Fermi operator of the form
$\epsilon_{\alpha\beta\gamma}u^{c\beta,T}_jCd^{c\gamma}_ku^{c\alpha,T}_iCe_l$.
(where $c$ in the superscript stands for charge conjugate).
This operator in the flavor basis must be antisymmetric in the exchange of
the two flavor indices $i$ and $j$. Once they are antisymmetric in the
flavor basis, they have to involve charm quark in the mass basis since
$uu$ terms will then be zero. Thus the to leading order the bino
contribution also vanishes.

The Higgsino exchange leads to an effective operator of the form:
\begin{equation}
I\epsilon_{\alpha\beta\gamma}(D_{kji'l'})X_{i'i,l'l}[u^{c\alpha,T}_k
Cd^{c\beta}_j(d^{\gamma,*}_iC\nu^*_l)+
u^{\beta,T}_jCd^{\gamma}_ku^{\alpha,T}_iCe_l]
\end{equation}
where $X_{i'i,l'l}\simeq \frac{1}{16\pi^2v \sin\beta
\cos\beta}M_{u,i'i}M_{\ell, l'l}$. Since $1/\sin \beta \cos\beta
\sim \tan\beta$ for large values of $\tan\beta$, this contribution
grows with $\tan\beta$. It is clear from inspection that the
largest value for this amplitude comes from $\widetilde{t}$
intermediate states and we estimate the largest contribution to be
of order $ C_{1323}\frac{m_t V_{ub}
m_{\tau}}{v^2_{wk}16\pi^2}\simeq 10^{-10}$ as compared to the LLLL
contribution which are of order $C_{1123}\frac{\alpha_2}{4\pi}
\sim 10^{-9}$. Therefore, we can ignore the RRRR contribution in
our discussion.

\section{Predictions for proton decay}
Let us first note that
the operators with $s$ quark lead to p-decay final states with $K$
meson whereas the ones without $s$ lead to $\pi$ final states. Also
generally speaking the amplitude for
nonstrange final states are down by a factor of Cabibbo angle $(\sim
0.22)$ compared to the strange final states as in the case of
SU(5) model. However, as we will see, we need to do a fine tuning among
the parameters $x,y,z$ to make the $p\rightarrow K^++\bar{\nu}$ compatible
with experiments. The same fine tuning however does not simultaneously
lower the amplitudes with nonstrange final states. As a result for some
domain of the allowed parameter space, one can have
the $p\rightarrow \pi^++\bar{\nu}$ mode as the dominant mode. This is very
different from the minimal SU(5) case.

In order to proceed to the calculation of proton lifetime, we must
extrapolate the above operators defined at the GUT scale first
to the $M_S$ and then to the one GeV scale. These extrapolation factor
have been calculated in the literature for MSSM and we take these values.
The required factors are: $A_L A_S$\cite{su50} and are given
numerically to be
$A_L=0.4$ (SUSY to one GeV scale) and $A_S=0.9-1.0$ (GUT to $M_{S}$
scale).

 The next step in the calculation is to go from three quarks to
proton. The parameter is denoted in the literature by $\beta$ and has
units of (GeV)$^3$. This has been calculated using lattice as well as
other methods and the number appears to be: $\beta\sim
0.007-0.028$\cite{beta}. We find that for our choice of the average
superpartner masses, for $\beta \geq 0.01$, there is
no range for the parameters $(x,y,z)$ where all decay modes have lifetimes
above the present lower limits. Of course as the superpartner masses
increase, larger $\beta$ values become acceptable. For instance, we note
that a change $\delta m^2_{\tilde{q}}/m^2_{\tilde{q}}$ by 10\% allows a
20\% higher value in $\beta$. We confine ourselves to the
domain $0.007 \leq (\beta/GeV^3) \leq 0.01$ and find that for all choices
of the free parameters allowed by the present lower limits, lifetimes
for the decay
modes $p\rightarrow \pi^+\bar{\nu}$ and $n\rightarrow \pi^0\bar{\nu}$ have
upper limits, which can therefore be used to test the model (see below).

Finally, in a detailed evaluation of proton decay rate to different final
states, we take into account the chiral symmetry
breaking effects following a chiral Lagrangian model (the first
two papers of Ref.\cite{beta}), where the chiral symmetry breaking effects
are parameterized by two parameters $D$ and $F$. These are usually chosen
to be the same as the analogous parameters in weak semileptonic
decays\cite{marshak}.

For this case, we find
the rate for proton decay to a particular decay mode $P\ell$ ($P$ is the
meson and $\ell$ denotes the lepton) to have the form:
\begin{eqnarray}
\Gamma_p(P\ell)~\simeq~\frac{m_p}{32\pi f^2_\pi M^2_T}|\beta|^2
\left(\frac{M_{\widetilde{W}}}{M^2_{\widetilde{f}}}\right)^2
\left(\frac{\alpha_2}{4\pi}\right)^2|A_LA_S|^24|C|^2|f(F,D)|^2\\ \nonumber
\simeq 2.7\times 10^{-50}|C|^2\left(\frac{2\times 10^{16}
GeV}{M_T}\right)^2\left(\frac{M_{\widetilde{W}}}{200~GeV}\right)^2
\left(\frac{TeV}{M_{\widetilde{f}}}\right)^4|f(F,D)|^2 ~ GeV
\end{eqnarray}
where $f(F,D)$ is a factor that depends on the hadronic parameters $F$ and
$D$ and we have used $\beta=0.01$ GeV$^3$ in the last expression.
We now discuss the evaluation of the parameter $|C|^2$ which determines
 the partial proton decay lifetimes for various modes. The relevant
modes are $p\rightarrow K^+\bar{\nu}$, $K^0\mu^+$, $K^0e^+$, $\pi
e^+$, $\pi\mu^+$. The present lower limits (including $n\rightarrow
\pi\nu, K\nu$ modes)
 on these modes are:
\begin{center}
{\bf Table II}

\begin{tabular}{|c||c|}\hline
mode & lifetimes ($\times 10^{32}$ yrs) \\ \hline
$p\rightarrow K^+\bar{\nu}$ &
$19$ \\ \hline $p\rightarrow K^0e^+$ &
$5.4$ \\ \hline $p\rightarrow K^0\mu^+$ & $10$ \\ \hline
$p\rightarrow \pi^+\bar{\nu}$
&
$0.2$\cite{kam}, $0.16$\cite{mann}
 \\ \hline $p\rightarrow \pi^0 e^+$ & $50$ \\ \hline
$p\rightarrow \pi^0\mu^+$ & $37$\\ \hline
$n\rightarrow \pi^0\bar{\nu}$ & $4.4$ \\ \hline
$n\rightarrow K^0\bar{\nu}$ & $1.8$\\ \hline
\end{tabular}
\end{center}

 {\bf Table caption:} Present experimental lower limits on the relevant
proton decay modes from Super-Kamiokande and Kamiokande experiments.

To proceed with this discussion, first note that $C$'s are products
of the known Yukawa coupling parameters $h$ and $f$ and the four GUT scale
parameters
as already discussed in Eq.(12). The GUT scale values of $h$ and $f$ are
obtained from neutrino fits described in Ref\cite{goh} and are given in
the previous section.

As far as the GUT scale parameters go, we will keep the overall
mass parameter to be the GUT scale i.e. $2\times 10^{16}$ GeV. We
have diagonalized the mass matrix of the color triplet GUT scale
Higgs fields in {\bf 10}, {\bf 126} etc and we find that they also
lead to the same parametrization as we have given here. The
meaning of the overall mass scale is then that it represents a
product of one of the mass eigenstates with the determinant. We
have checked that for the allowed range of parameters, the value
of the determinant, given by $|x-yz|$  is around 0.25 or so, so that
none of the mass eigenstates is too much higher than the GUT
scale. As a result the threshold effects on the gauge coupling
unification is minimal.

We then adopt the strategy
that we vary the parameters $x,y,z$ in such a way that the nucleon decay
rate to the $p\rightarrow K^+\bar{\nu}$ mode (summed over all the final
neutrino final states) is consistent with the present experimental lower
limit. Since there are three final states which add incoherently, this
narrows the space of the $x,y,z$ to a small domain. In this domain we
pick a point (call it $(x_0,y_0,z_0)$), where all other modes also satisfy
their present
experimental constraints as in Table II. We then vary the $(x,y,z)$
parameters around $(x_0,y_0,z_0)$ until the lifetime for a mode goes
below its present experimental lower limit. We find
that dependence on the parameter $z$ is much stronger than the others.
In Fig. 3 and 4, we give the allowed domain of the parameters $(x,y)$
consistent
with the various experimental lower limits on the partial lifetimes for
an optimum value of $z$. The boundary of the domain is determined by
the lower limit on the the $p\rightarrow K^+\bar{\nu}$. Inside this domain
the $\tau(p\rightarrow K^+\bar{\nu})$ is higher than its present lower
limit. The maximum value of the $p\rightarrow \pi^+\bar{\nu}$ and
$n\rightarrow \pi^0\bar{\nu}$ occurs at the boundary. We find that
$\tau(n\rightarrow \pi^0\bar{\nu})=~2\tau(p\rightarrow \pi^+\bar{\nu})$
has an upper bound of $ (5.7-13)\times 10^{32}$ yrs depending on whether
$\beta = 0.01-0.007$ GeV$^3$. At a different point in the parameter
space, $\tau(n\rightarrow K\bar{\nu})$ acquires its maximum value of
$2.9\times 10^{33}$ years. 
The predictions for the partial lifetimes of other modes are given in
Table III for both these cases. These values are accessible to the next
round of proton
decay searches.

\begin{center}

{\bf Table III}

\bigskip

\begin{tabular}{|c||c||c||c|}\hline
mode & $\tau/ 10^{32}$ yrs $\beta=0.01$ &$\tau/ 10^{32}$ yrs:
$\beta=0.007$ & $\tau/ 10^{32}$ yrs: $\beta=0.007$ \\
\hline 
    & $\tau(n\rightarrow \pi\bar{\nu})$ maximized &$\tau(n\rightarrow
\pi\bar{\nu})$ maximized & $\tau(n\rightarrow K\bar{\nu})$ maximized\\
\hline $p\rightarrow K^+\bar{\nu}$
&
$19$ & $19$ & $19$\\ \hline $p\rightarrow K^0e^+$ &
$1793$ & $2848$ & $188$\\ \hline $p\rightarrow K^0\mu^+$ & $184$ & $303$
& $28$ \\ \hline $p\rightarrow \pi^+\bar{\nu}$ &
$2.87$ & $6.5$ & $2.59$ \\ \hline $n\rightarrow \pi^0\bar{\nu}$ & $5.7$ &
$13$& $5.18$ \\ \hline $p\rightarrow \pi^0 e^+$ & $2452$ & $3857$ & $243$
\\ \hline $p\rightarrow \pi^0\mu^+$ & $263$
& $430$ & $37$ \\ \hline $n\rightarrow K^0\bar{\nu}$ & $1.9$ & $3.1$ &
$ 29.7$\\ \hline
\end{tabular}
\end{center}

\noindent {\bf Table caption}: Predictions for various nucleon decay modes
for the case when the lifetime for the mode $n\rightarrow \pi^0+\bar{\nu}$
attains its maximum value. The units for $\beta$ parameter
(i.e. GeV$^3$) has been omitted in the table. In column 4, we give the
lifetimes for the case when $\tau(n\rightarrow K\bar{\nu})$ is maximized.

We check the above results adopting an alternative strategy where we
express the three
parameters $(x,y,z)$ in terms of three partial life times and plot the
other lifetimes as a function of these partial life times. It turns out
that if we pick a certain value for the partial life time of the
$p\rightarrow K\mu$
mode and use it as an input, the other two input values get very
restricted. This allows us to use only the $p\rightarrow K^0\mu^+$ mode as
a variable and
give the others as a prediction.
In Fig. 5 and 6 we present the allowed values for various partial
lifetimes as a function of the partial lifetime for the mode $p\rightarrow
K^0\mu^+$. There is a slight spread around the various lines. We first
find that the lifetime for the mode $K^+\bar{\nu}$ can be arbitrarily
large as can be seen from Fig. 5. Also, from Fig. 5, we see that modes
$n\rightarrow \pi^0\bar{\nu}$ and $n\rightarrow K^0\bar{\nu}$ have upper
bounds which are same as the ones derived previously.

\begin{figure}[f]
\includegraphics[scale=0.70, bb= 30 0 523 400]{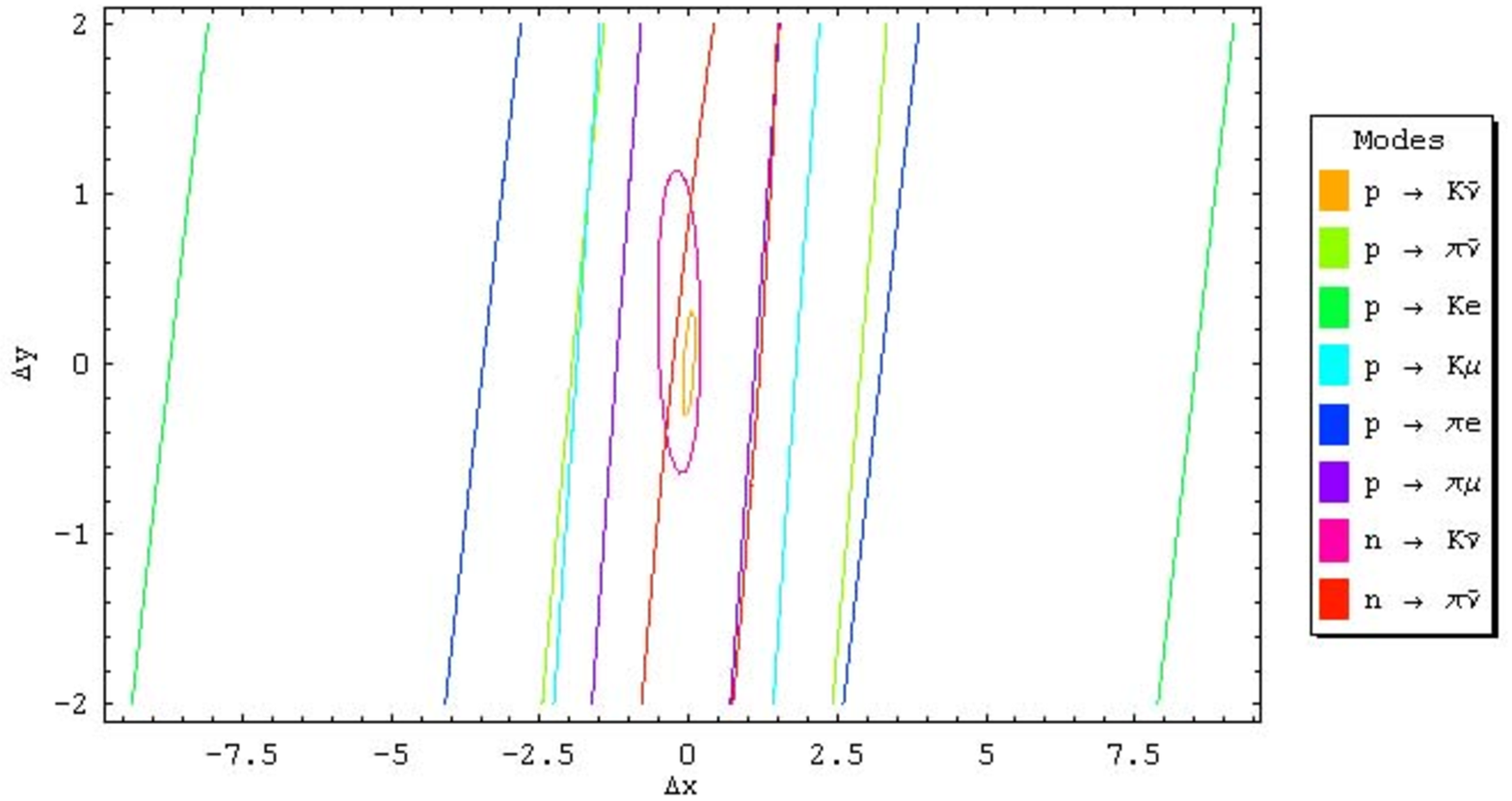} \caption{\label{fig3}
Allowed Region for $(x,y)$ coming from experimental lower limits
on lifetimes for different decay modes for $z=0.329$. The point
$(\Delta x,\Delta y)=(0,0)$ corresponds to $(x,y)=(-0.036,
0.387)$. Note that the region is most constrained by $p\rightarrow
K +\bar{\nu}$ mode.}
\end{figure}

\begin{figure}[f]
\includegraphics[scale=0.70, bb= 30 0 523 400]{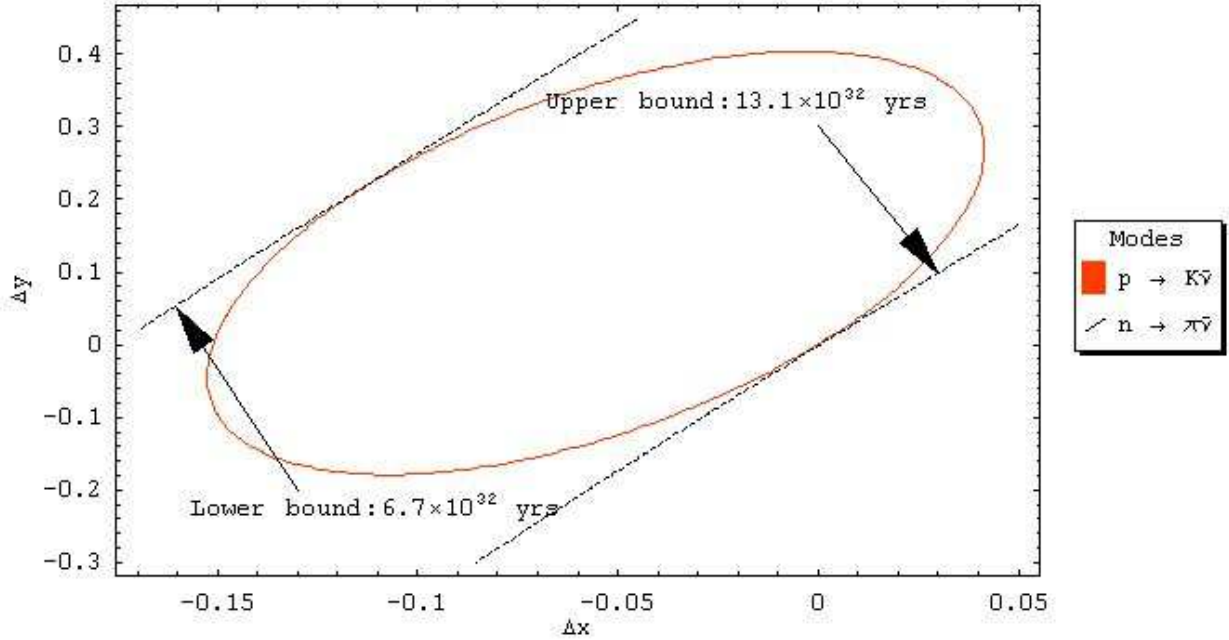}
\caption{\label{fig4}
Upper limit on the $n\rightarrow \pi +\bar{\nu}$ partial lifetime
while satisfying bounds on the lifetimes of all other modes. The
point $(\Delta x,\Delta y)=(0,0)$ corresponds to $(x,y,z)=(-0.132,
0.347,0.306)$.}
\end{figure}

\begin{figure}[f]
\includegraphics[scale=0.70, bb= 30 0 523 400]{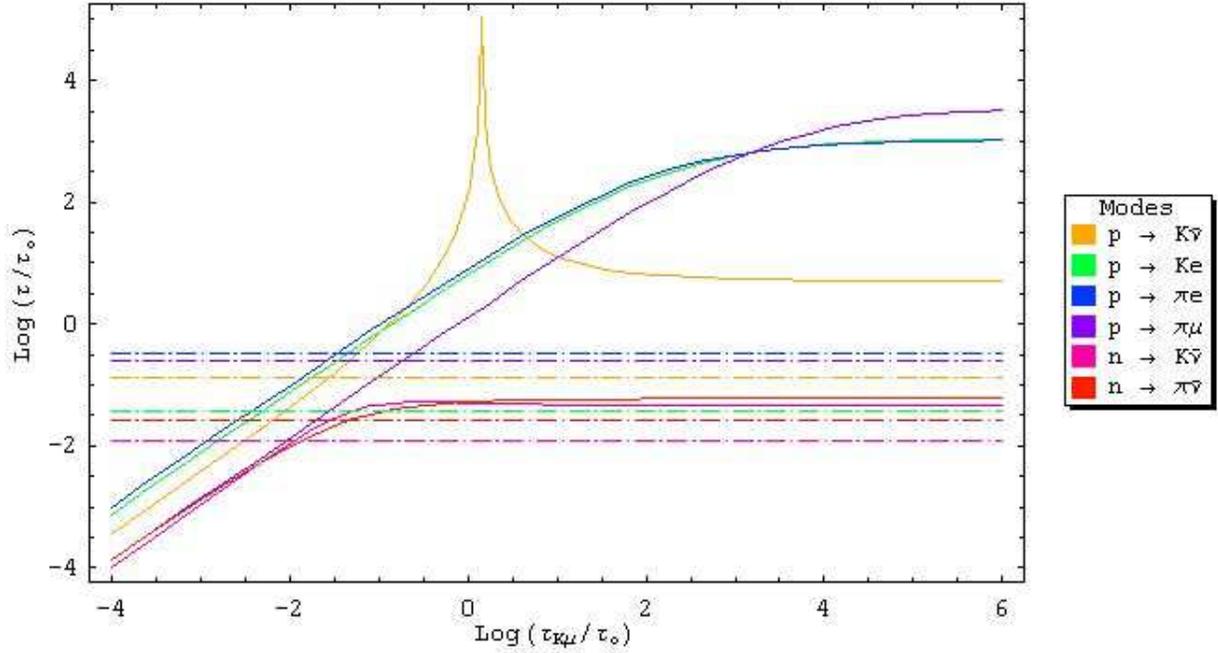} \caption{\label{fig7}
This figure gives the values of the lifetimes for different
proton decay modes as a function of the lifetime of the
$p\rightarrow K^0\mu^+$ mode (represented here by
$log_{10}\frac{\tau_{K\mu}}{\tau_0}$ where $\tau_0=14.6\times
10^{33}$ years) when $\tau(p\rightarrow K^+\bar{\nu})$ mode is at
its maximum value. This figure displays the values for one range of
$(x,y,z)$ and the following figure does it for a complementary
range. Also note that we have not included the gauge boson exchange
diagrams, which provide a value for these lifetimes around $10^{36}$
years or so.} \end{figure}

\begin{figure}[f]
\includegraphics[scale=0.70, bb= 30 0 523 400]{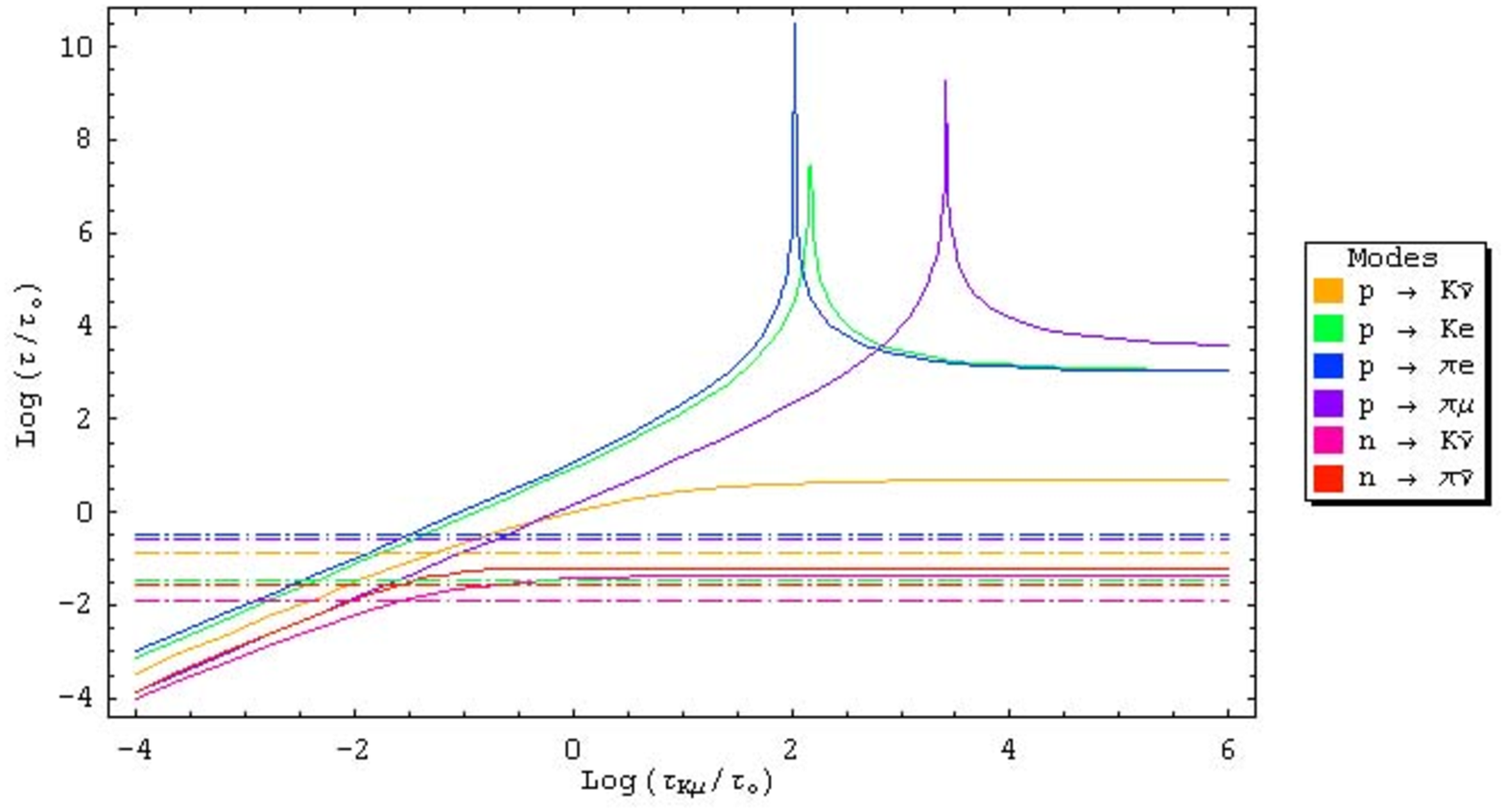} \caption{\label{fig8}
The same display as Fig. 5 but for a complementary range for the
parameters $(x,y,z)$.}
\end{figure}

\section{Conclusion}
 In summary, we have discussed the predictions for nucleon decay in the
minimal SO(10) model which has recently been shown to lead to
predictions for neutrino masses and mixings in agreement with
observations using only the assumption of type II seesaw
mechanism. The key feature of this model is that B-L symmetry is
broken by a single {\bf 126} field that also contributes to
fermion masses. For the range of the parameters that are allowed
by the neutrino data, we vary the GUT scale parameters ( unrelated
to the neutrino sector) so as to satisfy the stringent
experimental bounds for the decay mode $p\rightarrow
K^++\bar{\nu}$. We then predict an upper limit for the lifetimes
for the modes $p\rightarrow \pi^++\bar{\nu}$ and $n\rightarrow
\pi^0\bar{\nu}$ as follows: $\tau(n\rightarrow \pi^0\bar{\nu})=
2\tau(p\rightarrow \pi^+\bar{\nu})\leq 5.7-13\times 10^{32}$ years
and $\tau(n\rightarrow K\bar{\nu})\leq 2.9\times 10^{33}$ yrs for
the wino masses of 200 GeV and squark and slepton masses of a TeV.
This should provide new motivations for a new search for proton
decay, more specifically for these decay modes in question.

 This work is supported by the National Science Foundation
Grant No. PHY-0099544. We like to thank K. S. Babu, Tony Mann,
J. C. Pati, J. Schechter, G. Senjanovi\'c and  K. Turzynski for
useful discussions and comments.

\end{document}